\documentclass{llncs}
\usepackage[utf8]{inputenc}
\usepackage[T1]{fontenc}
\usepackage{courier}
\usepackage{stmaryrd}
\usepackage[clock]{ifsym}
\usepackage{pifont}
\usepackage{soul}
\usepackage[displaymath, mathlines]{lineno}
\usepackage[font=small]{caption}
\usepackage[font=small]{subcaption}
\usepackage{graphicx}
\usepackage{xspace}
\usepackage[inline]{enumitem}
\usepackage{alltt}
\usepackage{amsmath}
\usepackage{amssymb}
\usepackage{mathpartir}
\usepackage{mathtools}
\usepackage{hyperref}
\usepackage{why3lang}
\usepackage{simplebnf}
\usepackage{listings}
\usepackage{tikz}
\usetikzlibrary[calc]
\usepackage[capitalise]{cleveref}
\usepackage{xcolor}

\usepackage{cleveref}
\newcommand{\eg}{\textit{e.g.}\xspace}
\newcommand{\ie}{\textit{i.e.}\xspace}

\newcommand{\tool}[1]{\textsf{#1}\xspace}

\newcommand{\yyy}{\tool{Why3}}
\newcommand{\caisar}{\tool{CAISAR}}
\newcommand{\vehicle}{\tool{Vehicle}}
\newcommand{\marabou}{\tool{Marabou}}
\newcommand{\maraboupy}{\tool{maraboupy}}
\newcommand{\pyrat}{\tool{PyRAT}}
\newcommand{\nnenum}{\tool{nnenum}}
\newcommand{\abcrown}{\tool{$\alpha$-$\beta$-CROWN}}
\newcommand{\saver}{\tool{SAVer}}
\newcommand{\vnnl}{\tool{VNN-LIB}}
\newcommand{\smtl}{\tool{SMT-LIB}}
\newcommand{\vnnc}{\tool{VNN-Comp}}
\newcommand{\onnx}{\tool{ONNX}}
\newcommand{\nnet}{\tool{NNet}}
\newcommand{\qflra}{\tool{QF\_LRA}}

\newcommand{\yml}{\tool{WhyML}}
\newcommand{\nir}{\tool{NIR}}
\newcommand{\aimos}{\tool{AIMOS}}
\newcommand{\acas}{\tool{ACAS Xu}}

\newcommand{\cmark}{\ding{51}}%
\newcommand{\xmark}{\ding{55}}%

\newcommand{\kw}[1]{\textbf{\texttt{#1}}}
\newcommand{\readm}[1]{\ensuremath{\textnormal{\texttt{read\_model}}\;\mathtt{#1}}}

\newcommand{\vlen}[1]{\ensuremath{\textnormal{\texttt{length}}\;#1}}
\newcommand{\hlen}[2]{\ensuremath{\textnormal{\texttt{has\_length}}\;#1\;#2}}
\newcommand{\atat}[2]{\ensuremath{#1\,@@\,#2}}
\newcommand{\vget}[2]{\ensuremath{#1}\ensuremath{\left[\,#2\,\right]}}

\newcommand{\fall}[2]{\ensuremath{\textnormal{\texttt{forall\_}}\;#1\;#2}}
\newcommand{\mapyyytonir}[1]{\llbracket#1\rrbracket}
\newcommand{\nonterm}[1]{\ensuremath{\langle \mathit{#1} \rangle}}

\makeatletter
\newcommand*{\rectxy@anchor@top}[2]{%
  \anchor{t#1}{%
    \pgf@process{\southwest}%
    \pgf@xa=\pgf@x
    \pgf@process{\northeast}%
    \pgf@x=\dimexpr\pgf@xa + (\pgf@x-\pgf@xa)*#1/#2\relax
  }%
}
\newcommand*{\rectxy@anchor@bottom}[2]{%
  \anchor{b#1}{%
    \pgf@process{\northeast}%
    \pgf@xa=\pgf@x
    \pgf@process{\southwest}%
    \pgf@x=\dimexpr\pgf@x + (\pgf@xa-\pgf@x)*#1/#2\relax
  }%
}
\newcommand*{\rectxy@anchor@left}[2]{%
  \anchor{l#1}{%
    \pgf@process{\northeast}%
    \pgf@ya=\pgf@y
    \pgf@process{\southwest}%
    \pgf@y=\dimexpr\pgf@y + (\pgf@ya-\pgf@y)*#1/#2\relax
  }%
}
\newcommand*{\rectxy@anchor@right}[2]{%
  \anchor{r#1}{%
    \pgf@process{\southwest}%
    \pgf@ya=\pgf@y
    \pgf@process{\northeast}%
    \pgf@y=\dimexpr\pgf@ya + (\pgf@y-\pgf@ya)*#1/#2\relax
  }%
}
\newcommand*{\declareshaperectxy}[4]{%
  \pgfdeclareshape{rectangle #1x#2x#3x#4}{%
    \inheritsavedanchors[from=rectangle]
    \inheritanchorborder[from=rectangle]
    \inheritanchor[from=rectangle]{north}
    \inheritanchor[from=rectangle]{north west}
    \inheritanchor[from=rectangle]{center}
    \inheritanchor[from=rectangle]{west}
    \inheritanchor[from=rectangle]{east}
    \inheritanchor[from=rectangle]{mid}
    \inheritanchor[from=rectangle]{mid west}
    \inheritanchor[from=rectangle]{mid east}
    \inheritanchor[from=rectangle]{base}
    \inheritanchor[from=rectangle]{base west}
    \inheritanchor[from=rectangle]{base east}
    \inheritanchor[from=rectangle]{south}
    \inheritanchor[from=rectangle]{south east}
    \inheritbackgroundpath[from=rectangle]
    \count@=\m@ne
    \@whilenum\count@<#1 \do{%
      \advance\count@\@ne
      \expandafter\rectxy@anchor@top\expandafter{\the\count@}{#1}%
    }%
    \count@=\m@ne
    \@whilenum\count@<#2 \do{%
      \advance\count@\@ne
      \expandafter\rectxy@anchor@bottom\expandafter{\the\count@}{#2}%
    }%
    \count@=\m@ne
    \@whilenum\count@<#3 \do{%
      \advance\count@\@ne
      \expandafter\rectxy@anchor@left\expandafter{\the\count@}{#3}%
    }%
    \count@=\m@ne
    \@whilenum\count@<#4 \do{%
      \advance\count@\@ne
      \expandafter\rectxy@anchor@right\expandafter{\the\count@}{#4}%
    }%
  }%
}
\makeatother

\declareshaperectxy{1}{1}{3}{1}
\usepackage{array}
\begin{document}
\title{The \caisar Platform: Extending the Reach of Machine Learning Specification and Verification}
\titlerunning{The \caisar Platform}
%
\author{
  Michele Alberti\textsuperscript{\href{https://orcid.org/0009-0001-4352-4555}{\includegraphics[height=5pt]{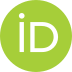}}}
  \and
  François
  Bobot\textsuperscript{\href{https://orcid.org/0000-0002-6756-0788}{\includegraphics[height=5pt]{orcid_logo.png}}}
  \and
  Julien Girard-Satabin\textsuperscript{\href{https://orcid.org/0000-0001-6374-3694}{\includegraphics[height=5pt]{orcid_logo.png}}}
  \and
  Alban Grastien\textsuperscript{\href{https://orcid.org/0000-0001-8466-8777}{\includegraphics[height=5pt]{orcid_logo.png}}}
  \and
  Aymeric Varasse\textsuperscript{\href{https://orcid.org/0009-0007-0829-9309}{\includegraphics[height=5pt]{orcid_logo.png}}}
  \and
  Zakaria Chihani\textsuperscript{\href{https://orcid.org/0009-0004-8915-4774}{\includegraphics[height=5pt]{orcid_logo.png}}}
}
\authorrunning{M. Alberti, F. Bobot, J. Girard-Satabin, A. Grastien, A. Varasse, Z. Chihani}
%
\institute{Université Paris-Saclay, CEA, List, F-91120, Palaiseau, France
  \email{firstname.lastname@cea.fr}\\ \email{julien.girard2@cea.fr} \\
  Software Heritage \\ \email{aymeric.varasse@softwareheritage.org}}
\maketitle              
\begin{abstract}
  The formal specification and verification of machine learning models have
  advanced remarkably in less than a decade, leading to a profusion of
  verification tools that provide mathematical guarantees about model
  properties.
  However, this growing diversity risks ecosystem fragmentation, making it
  difficult to compare tools beyond narrowly defined benchmarks.
  Moreover, much of the progress to date has focused on a limited class of
  properties, particularly \emph{local robustness}.
  While existing tools are increasingly effective at verifying such properties,
  more complex ones, such as those involving multiple neural networks, remain
  beyond their capabilities: these properties cannot currently be expressed in
  their specification languages, nor can they be directly verified.
  This applies even to the winning verification tools of the International
  Verification of Neural Networks Competition (\vnnc).

  In this tool paper, we present \caisar, an open-source platform for specifying
  and verifying properties of machine learning models, with particular focus
  on neural networks and support vector machines.
  \caisar provides a high-level language for specifying complex properties and
  integrates several state-of-the-art verifiers for their automatic
  verification.
  Through concrete use cases, we show how \caisar leverages automated
  graph-editing techniques to translate high-level specifications into queries
  for the supported verifiers, bridging the (embedding) gap between user
  specifications and the corresponding ones that are actually verified.
  %

  \keywords{Formal Specification and Verification, Machine Learning}
\end{abstract}
\section{Introduction}
\label{sec:intro}
In recent years, the formal methods (FM) community has made significant
progress, going from the very first neural network (NN)
verifiers~\cite{pulina2011never,katz2017reluplex}, suffering from scalability
issues, to a wide range of tools capable of handling increasingly large
NNs~\cite{katz2019marabou,wang2021beta,bak2021nnenum,durand2022reciph,lemesle2024pyrat,singh2019single}\footnote{For
  a comprehensive overview of the field, we refer readers to the survey
  in~\cite{urban2021review}.} and diverse machine learning (ML) models, like
Support Vector Machines (SVM)~\cite{saver,cristianini2008support} and boosted
trees~\cite{audemard2024computation}.

From this Cambrian explosion of tools, we draw the following observations:

\paragraph{A focus on a narrow set of properties}
Research on the formal verification of ML models has largely concentrated on a
restricted class of properties, most notably \emph{local robustness}.
Given an input $x$, a model $f$, and a perturbation bound $\epsilon \in
  \mathbb{R}$, local robustness is defined as the requirement that
\[
  \forall x'.\ \|x - x'\| \leq \epsilon \implies f(x) = f(x').
\]
Several variants of this definition have been proposed~\cite{casadio2022neural},
including \emph{$K$-Lipschitz robustness}~\cite{balan2017lipschitz}
and \emph{strong classification robustness}~\cite{fischer2019dl2}.
The emphasis on local robustness is motivated by the nature of modern ML models,
which frequently operate on high-dimensional inputs such as images or text.
In such settings, formulating functional specifications is often
impractical---if not impossible---whereas local robustness remains applicable
and well-defined.
While this has not prevented researchers from proposing higher-level
specifications---including fairness constraints~\cite{urban2019perfectly},
semantic transformations robustness~\cite{balunovic2019certifying} or providing
formally-grounded explanations~\cite{bassan2023towards,wu2023verix}---we argue
that the field remains primarily centered on robustness properties.
As previously noted~\cite{cordeiro2025neural}, there is an urgent need for more
expressive specification languages that would enable the formal verification of
a broader range of semantic properties of ML models.

\paragraph{Limitations on the expressiveness of input languages}
\label{subsec:intro_vnnlib}

The current de-facto standard for specifying properties is the \vnnl language,
adopted by the Verification of Neural Network Competition
(\vnnc)~\cite{brix2023first,brix2023fourth,brix2024fifth}.
As the \vnnl was originally designed to handle the community focus towards the
specification and verification of \emph{local robustness} properties of NNs, it
exhibits certain limitations. First, the size of specification increases with
the size of the NN, making it cumbersome for a human to write, and even more so
to read and maintain. Second, \vnnl restricts specifications to conjunctions of
linear arithmetic constraints over inputs and outputs, which are the only
authorized free variables.

More formally, let $x \in \mathbb{R}^d$ be a vector and $x_c$ its $c$-th
element, $f : \mathbb{R}^d \mapsto \mathbb{R}^p$ be a function representing an
NN that takes a $d$-dimensional vector as input and returns a $p$-dimensional
vector, $\vec{K} \subseteq \{1,\dots,p\}$ be a (possibly empty) set of integers,
and $a,b \in \mathbb{R}^d$ and $c,d \in \mathbb{R}^p$ be vectors. A \vnnl
specification is then equivalent to the following formula:
\begin{equation*}
  \begin{aligned}
    \forall x \in \mathbb{R}^d.\
    \forall y \in \mathbb{R}^p.\
    y = f(x) \implies                                     &
    \\
    \bigwedge_{i = 0 .. d} a_i \leq x_i \leq b_i \implies &
    (\bigwedge_{j =
      0 .. p} c_j \leq y_j \leq d_j
    \land
    \bigwedge_{k \in \vec{K}} y_p \leq y_k),
  \end{aligned}
\end{equation*}
which specifies that if the input is within a polyhedron defined by $a$ and $b$,
then the output should be within the polyhedron defined by $c$ and $d$, and its
$p$-th element should be lower than all others in $\vec{K}$.

Such a formulation prevents expressing properties on multiple
NNs~\cite{athavale2024verifying},
hyperproperties~\cite{xie2022neuro,boetius2023verifying}, or any specification
yielding uninterpreted functions. Promising high-level specification languages
exist, notably the one in \vehicle~\cite{daggitt2022vehicle}, that
need to be both expressive enough \emph{and} compile down to a vast range of
verifiers to take advantage of the profusion of existing tools.

\paragraph{Difficulty to compare and select a verifier}
A profusion of solutions implies the burden of choice. For now, this burden is
mostly taken by the end-user of verification tools. Comparing the capabilities of
tools is a first prerequisite to lift that burden. This need for comparison
is partially addressed with the \vnnc{}, where
tools are evaluated against benchmarks proposed by the community.
However, as displayed in~\cite{koenig2024critically}, this is but one
of the numerous facets for evaluating the quality of verifiers.
Indeed, depending on the
property to check or the implementation specifics of a ML model, some verifiers
may not be suitable, or underperform compared to
others~\cite{koenig2024critically},
making it difficult to assess the strengths and weaknesses of
specific verifiers. It is thus necessary to provide a systematic and principled
way to compare verifiers against the same basis.

\paragraph{A gap between the specified and proven programs}
Formal verification is a multifaceted activity that usually involve multiple
steps: writing a specification, developing a program, encoding the
specification into a suitable tool and ensuring the program is
correct regarding the encoded specification.

A common struggle is to ensure that each
step of the process is trustworthy and coherent one with another, that is to
say, that the proof run on a particular program is meaningful with regard to a
specification that was provided.
One possible solution is to couple writing the
specification \emph{and} generating the code, like in the industrial software
\tool{SCADE}\footnote{\url{https://www.ansys.com/en-gb/products/embedded-software/ansys-scade-suite}}.
However, machine learning has its own peculiarities. For instance, it is common
to apply transformations on the inputs of a neural network during the evaluation
pipeline to normalize inputs regarding a certain training
set distribution. Such transformations are not specified inside the NN, but
done at runtime. Furthermore, most tools express specifications with real
arithmetic, whereas NNs
compute floating points numbers.
\vnnl assumes that the companion
program to verify is a NN, but without specifying anything towards
it, or the transformations on inputs.
This creates an \emph{embedding
  gap}~\cite{daggitt2024vehicle,cordeiro2025neural}: the specification the program is checked against
follows an unspecified transformation.
Ideally, a complete specification should be unambiguous on the transformations
applied to an input.\newline

In this paper, we present a way to lift these limitations within the open-source
verification platform \caisar~\cite{alberti2022caisar}.
We present and implement a higher-level specification language that allows users
to express complex properties, including properties with multiple ML models.
\caisar proposes an automated and principled way to translate specifications written in a
higher-level language to 9 verification tools---including the winners of the
\vnnc{}.
\caisar is thus able to express properties and send them to existing
off-the-shelf verification tools
\emph{without modifying them}. Its maturity allowed it to be used in
industrial settings on non-public use-cases, although this paper focuses on illustrating the platform through simple examples (see \cref{sec:eval}).

\paragraph{Artifact Availability Statement}
An artifact---including instructions, datasets, and experiment
scripts---reproducing the results in this paper is publicly available at
\url{https://doi.org/10.5281/zenodo.16902886}.

%
%
\section{General Architecture and Core Components}
\label{sec:architecture}
\begin{figure}[t]
  \begin{center}
    \includegraphics[width=\textwidth]{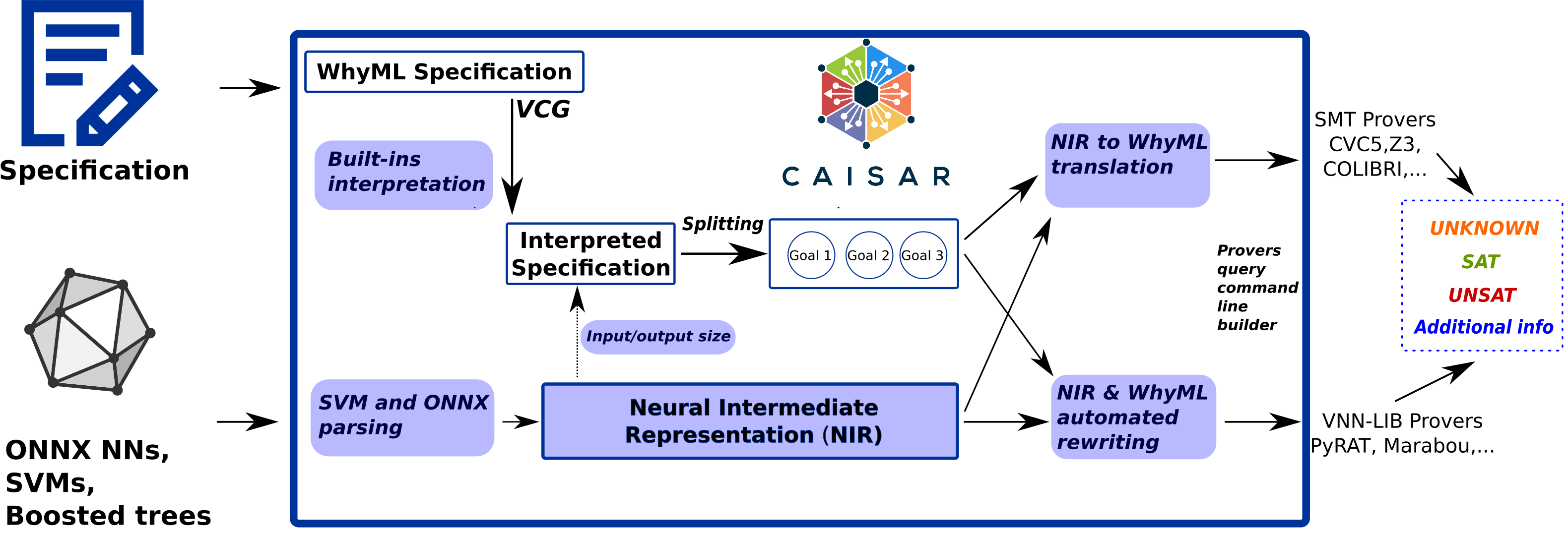}
  \end{center}
  \caption{The overall functional architecture of \caisar. Significant
    extensions to the original \yyy platform are outlined in shaded
    rectangles.}
  \label{fig:caisar_archi}
\end{figure}

The general architecture of \caisar is depicted in \cref{fig:caisar_archi}.
\caisar builds upon the design of \yyy~\cite{filliatre2013why3} and makes
extensive use of the \yyy API.
\yyy is a mature platform for deductive program verification, applied across a
wide range of domains, from robotic navigation
algorithms~\cite{trojanek2014verification} to the formalization and verification
of quantum programs~\cite{chareton2021automated}. It provides a rich input
language for specifying, implementing, and formally verifying the correctness of
algorithms.
Verification proceeds by generating verification conditions, which are
dispatched to a variety of automated and interactive provers via composable
transformations. A dedicated verification condition generator (VCG) translates a
program together with its specifications into logical goals. Finally, \yyy
offers a unified text-based configuration mechanism for each supported solver.

\caisar adapts those components to the specific needs of verification of ML models.
To include ML into the specification and ease
vector computation, we extend the \yyy input language with several \emph{built-ins}
as described in~\cref{sec:speclang}.
To handle those extensions, we modify the internal \yyy interpretation engine
as described in~\cref{sec:interpretation}.
To represent ML components in \yyy, we build a Neural Intermediate Representation
(\nir) described in~\cref{sec:nir}; \cref{sec:nnsvm} describes
how to translate SVMs to valid \nir representations.
High-level specifications are currently not directly amenable for provers.
In such cases, we translate part of the \yyy specification directly
as a new \nir as explained in~\cref{sec:merging}, using automated graph editing
that incorporates part of \yyy terms within the \nir.
Finally, a list of supported provers is presented in~\cref{sec:provers}.

\subsection{The Specification Language}
\label{sec:speclang}
\caisar supports specifications written in
\yml\footnote{\url{https://www.why3.org/doc/syntaxref.html}}, the specification
and programming language of \yyy.
It exhibits three key qualities: it is \emph{strongly typed} to prevent type
errors, \emph{expressive} to specify properties beyond local robustness, and
\emph{high-level} to bridge the embedding gap inherent of \vnnl (see
Section~\ref{sec:intro}).
Specifically, \yml enables writing properties of (multiple) ML models, treating
them as \emph{pure} (\ie side-effect free), \emph{abstract} functions that
operate on vectors.

To support this abstraction, \caisar extends the \yyy standard library with
dedicated \emph{theories}, such as the theory \texttt{Model} of ML models, and
\texttt{Vector} of finite-length arrays.
These theories define types, predicates and functions necessary to specify
properties about ML models, vectors and their operations.

We present the grammar of the language in \figurename\ref{fig:grammar}.
A \caisar specification is then a \yml sequence of type, predicate and function
definitions.

\begin{figure}[htb]
  \begin{minipage}[t]{0.49\textwidth}
    \strut\vspace*{-\baselineskip}\newline
    \begin{bnf}(prod-sep = {5pt},)[colspec = {lrcll},colsep=2pt]
      \nonterm{decl} ::=
      | \kw{type} \nonterm{tId} = \nonterm{type}
      | \kw{predicate} \nonterm{id} \\ &&& $\quad$ \nonterm{binder}$^*$ = \nonterm{expr}
      | \kw{function} \nonterm{id} \\ &&& $\quad$\nonterm{binder}$^*$ \nonterm{spec}$^*$ = \nonterm{expr}
      ;;
      \nonterm{type} ::=
      | \nonterm{tId}
      | \nonterm{type} $\rightarrow$ \nonterm{type}
      | (\nonterm{type},$\dots$,\nonterm{type})
      | \texttt{vector} \nonterm{type}
      | \texttt{int} // \texttt{bool} // \texttt{float} // \texttt{string}
      | \texttt{model}
      ;;
      \nonterm{binder} ::=
      | \nonterm{id} // (\nonterm{id}$\colon$\nonterm{type})
      ;;
      \nonterm{spec} ::=
      | \kw{requires} $\left\{\textnormal{\nonterm{expr}}\right\}$
      | \kw{ensures} $\left\{\textnormal{\nonterm{expr}}\right\}$
      ;;
      \nonterm{bop} ::=
      | $\leq$ // $\geq$ // $<$ // $>$
      | $+$ // $-$ // $\times$ // $/$
      | $\wedge$ // $\vee$ // $\rightarrow$
    \end{bnf}
  \end{minipage}%
  %
  \begin{minipage}[t]{.51\textwidth}
    \strut\vspace*{-\baselineskip}\newline
    \begin{bnf}(prod-sep = {5pt},)[colspec = {lrcll},colsep=2pt]
      \nonterm{expr} ::=
      | \nonterm{id}
      | \nonterm{built\textnormal{-}in}
      | \nonterm{expr}\nonterm{expr}
      | (\nonterm{expr},$\dots$,\nonterm{expr})
      | \kw{let} \nonterm{id} $=$ \nonterm{expr} \kw{in}
      | \kw{if} \nonterm{expr} \kw{then} \nonterm{expr}\\ &&&\kw{else} \nonterm{expr}
      | \nonterm{expr}\nonterm{bop}\nonterm{expr}
      | \kw{forall}\nonterm{binder}$.$\nonterm{expr}
      | \kw{exists}\nonterm{binder}$.$\nonterm{expr}
      | \kw{not}\nonterm{expr}
      | $\mathtt{i}\in\mathtt{Integer}$
      | $\{\mathtt{true},\mathtt{false}\}\in\mathtt{Boolean}$
      | $\mathtt{f}\in\mathtt{Float}$ // $\mathtt{s}\in\mathtt{String}$
      ;;
      \nonterm{built\textnormal{-}in} ::=
      | \texttt{read\_model} \nonterm{expr}
      | \texttt{length} \nonterm{expr}
      | \texttt{has\_length} \nonterm{expr} \nonterm{expr}
      | \nonterm{expr}$\left[\textnormal{\nonterm{expr}}\right]$ | \nonterm{expr}$@@$\nonterm{expr}
    \end{bnf}
    \vspace{-0.2cm}
  \end{minipage}
  \caption{Grammar of \caisar's high-level specification language. The
    nonterminal \nonterm{built\textnormal{-}in} denotes \yml functions having a
    special interpretation in \caisar.}
  \label{fig:grammar}
\end{figure}
%
%

%


Examples of specifications are shown in
\figurename\ref{fig:specexamplesequencing} and \figurename\ref{fig:specexample}.
\figurename\ref{fig:specexamplesequencing} shows a specification where the
output of a first NN is perturbed and then passed as an input to a second NN,
requiring the classification of the second NN to be a certain class.
This example illustrates how to naturally express properties involving multiples
NN, intermediate computations on inputs and outputs, and NN composition.
On the other hand, property in \figurename\ref{fig:specexample} specifies that
two NNs produce output vectors that differ by at most $\delta$ over their
entire, valid input domain.

\begin{figure}[t!]
  \centering
  \begin{lstlisting}[numbers=left]
  goal sequencing:
    let nn1 = Model.read_model "path/to/nn1.onnx" in
    let nn2 = Model.read_model "path/to/nn2.onnx" in
    let dataset = Dataset.read_dataset "path/to/csv" in
    let eps = (0.01:t) in
    CSV.forall_ dataset (fun l e ->
      forall perturbed_e.
        has_length perturbed_e (length e) ->
        FeatureVector.valid feature_bounds perturbed_e ->
        let perturbation = perturbed_e - e in
        ClassRobustVector.bounded_by_epsilon perturbation eps ->
        let out_1 = nn_1@@perturbed_e in
        let out_2 = nn_2@@out_1 in
        forall j. Label.valid label_bounds j -> j <> l ->
        out_2[l] .>= out_2[j]
    )
\end{lstlisting}
  \caption{\caisar specification about the output prediction of a composition
    of NNs.
    The specification begins by declaring a verification goal
    named \texttt{sequencing}.
    Lines 2-4 introduce two NNs and the dataset by loading their \onnx model
    and CSV files, followed by line 5 which sets the perturbation $\epsilon$
    of the first input to $0.01$.
    Line 6 ensures the specification will be checked against all samples in the
    dataset.
    Lines 7-11 define a perturbed vector that has the same length of the original
    vector and have the same bounded values.
    Lines 12-13 describe the computation: the input of the first NN is perturbed,
    then the computation is performed, and the composition by the second NN takes
    place.
    Finally, line 14-15 assert that the prediction for the given output label is
    always the preferred one.
  }
  \label{fig:specexamplesequencing}
\end{figure}

%
%

\begin{figure}[t!]
  \centering
  \begin{lstlisting}[numbers=left]
  goal equality_up_to_delta:
    let nn1 = Model.read_model "path/to/nn1.onnx" in
    let nn2 = Model.read_model "path/to/nn2.onnx" in
    let delta = (0.125:t) in
    forall input. Vector.has_length input 5 ->
      (forall i. 0 <= i < Vector.length input -> 0.0 .<= input[i] .<= 1.0) ->
        let output1 = nn1 @@ input in
        let output2 = nn2 @@ input in
        .- delta .<= output1[0] .- output2[0] .<= delta

\end{lstlisting}
  \caption{\caisar specification about the output equality up-to $\delta$
    between two NNs.
    The specification begins by declaring a verification goal named
    \texttt{equality\_up\_to\_delta}.
    Lines 2-3 introduce two NNs by loading their \onnx model files, followed by line
    4 which sets the tolerance threshold $\delta$ to $0.125$.
    Lines 5-6 constraint the input vector to be of length 5 with elements in $[ 0.0,
          1.0 ]$.
    Lines 7-8 then apply both NNs to this input using the \texttt{@@} operator
    for model application.
    Finally, line 9 asserts that the absolute difference between the first
    output components of both networks is bounded by $\delta$. }
  \label{fig:specexample}
\end{figure}

The \emph{built-ins} currently have no explicit definition inside \yml; rather
their semantics is determined via interpretation, a process described in more
detail in Section~\ref{sec:interpretation}.
For instance, the \texttt{Model} theory provides the built-ins \readm{s}, which
introduces a model in a specification by reading the filename $\mathtt{s}$, and
\atat{m}{v}, which returns a (output) vector obtained by applying model $m$ to
(input) vector $v$.
Similarly, the \texttt{Vector} theory provides several built-ins such as, among
others, \vlen{v} to return the length of vector $v$, \hlen{v}{i} to check
whether vector $v$ has length $i$, \vget{v}{i} to retrieve the element at
position $i$ of vector $v$, and \fall{v}{f} to verify that predicate $f$ holds
for every element of vector $v$.
%



%
\subsection{The Neural Intermediate Representation}
\label{sec:nir}
\subsubsection{Preliminaries on \onnx{}}

\begin{figure}
  \centering
  \includegraphics[width=\textwidth]{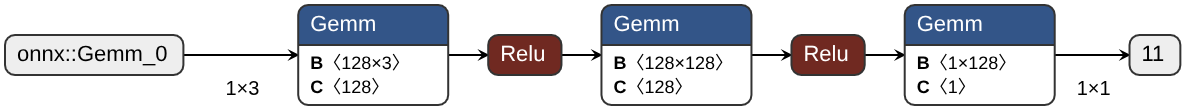}
  \caption{Example of an \onnx graph, obtained using the \tool{netron} tool.
    \kw{Gemm} stands for GEneralized Matrix Multiplication, while \kw{Relu} is
    the Rectified Linear Unit function.}
  \label{fig:example_onnx}
\end{figure}

The Open Neural Network Exchange (\onnx) is an intermediate representation
described via the \emph{Protocol Buffer}\footnote{\url{https://protobuf.dev/}}
interface language. It is vastly supported within state-of-the-art machine
learning frameworks like \tool{PyTorch} or \tool{TensorFlow}. \onnx{}
describes a directed acyclic graph (DAG) whose nodes are operators describing
computations on multidimensional arrays, called \emph{tensors}, and edges
describe the flow of data. An example of \onnx graph is available in
\figurename\ref{fig:example_onnx}.

At the time of writing, \onnx{} defines 193
operators\footnote{\url{https://github.com/onnx/onnx/blob/main/docs/Operators.md}}.
Below, we provide an informal semantics for a subset of these operators used later in the paper:
\begin{itemize}
  \item \kw{Add}$(x, y)$ (respectively, \kw{Sub}, \kw{Mul}, \kw{Div}) performs
        element-wise addition (respectively, subtraction, multiplication, and
        division) of tensors $x$ and $y$, which are assumed to have the same shape;
  \item \kw{Concat}$(x, y)$ appends tensor $y$ to tensor $x$ along a specified
        axis;
  \item \kw{Gather}$(x, y)$ selects values from tensor $x$ at indices specified
        by tensor $y$, that is, \kw{Gather}$(x, y) = x_{y_i}$;
  \item \kw{Gemm}$(x, y)$ performs matrix multiplication of tensors $x$ and $y$,
        optionally followed by the addition of a bias term $c$;
  \item \kw{Sign}$(x)$ returns an integer tensor $y$ such that $y_i =
         1$ whenever $x_i > 0$, $y_i = -1$ whenever $x_i < 0$, and $y_i = 0$ otherwise;
  \item \kw{ReLU}$(x)$ applies to tensor $x$ the rectified linear function
        $\max(0, x)$ element-wise;
  \item \kw{Input} designates the entry point of the graph\footnote{In plain
          \onnx, the entry point is specified via graph metadata. We define an
          operator \kw{Input} explicitly here for clarity.}.
\end{itemize}

\subsubsection{Representing Machine Learning Components}

\caisar defines a Neural Intermediate Representation (\nir) to model machine
learning components.
Similar to \onnx, a \nir is a DAG, where each node
encodes an operation, its input and output shapes, and the computation type. The
\nir language supports a subset of the \onnx Intermediate Representation v8
and \onnx Opset v13
standards\footnote{\url{https://onnx.ai/onnx/operators/index.html}}.
The full list of supported operators is available in the
\href{https://git.frama-c.com/pub/caisar/-/blob/master/lib/nir/node.mli?ref_type=heads}{\caisar
  source code}.

\caisar can parse various machine learning models into \nir, including neural
networks in the \onnx and \nnet formats, and support vector machines exported in
a format compliant with the \tool{scikit-learn}~\cite{walt2014scikit} framework.
Gradient boosting machines are currently handled via a separate pipeline.

The \nir is implemented as a pure recursive \tool{OCaml} variant type.
Its signature is compatible with the \tool{ocamlgraph}~\cite{conchon2007designing}
library, enabling efficient graph traversal and transformation.

\caisar also supports back-translation from \nir into multiple target formats.
It can generate valid \onnx models, and leverages prior
work~\cite{girard-satabin2020camus} to compile control-flow representations of
neural networks into the \smtl (and its subset, \vnnl) format, enabling
interoperability with a range of off-the-shelf solvers.

\subsubsection{Automated Transformations of SVM to \nir}
\label{sec:transformations2}
\caisar{} is able to reason about Support Vector Machines (SVM) by translating
them into equivalent \nir.

SVMs are classical ML models used for classification. Given an input vector $x$
and assuming $m$ classes, the model outputs a vector $y \in \mathbb{R}^m$ where
each component $y_{cl}$ is the \emph{score} associated with the class $cl \in
  \{1,\dots,m\}$.

We first consider the binary case ($m = 2$). In this setting, training
identifies a subset of the samples, called \emph{support vectors}, that lie on
or near the margin of separation between the two classes.
Together with their learned coefficients, these vectors define a hyperplane that
partitions the input space into two regions, one for class $cl_1$ and the other
for class $cl_2$.%
\footnote{This geometric interpretation is valid for the linear case; for models
  using Radial Basis Function (RBF) kernels, the separation is non-linear, which
  we omit here for simplicity.}

Multiple support vectors can contribute to the separation of the two classes.
Each class $cl_j$ has a (possibly different) number $k_j$ of support vectors,
denoted $S_{cl_j,1}, \dots, S_{cl_j,k_j}$.
Classification is based on the (signed) distance between the input $x$ and each
hyperplane defined by the support vectors, denoted $x \otimes S_{cl_j,\ell}$,
indicating how \emph{confidently} the corresponding support vector classifies
the input.

An optional non-linear function $f$ can be applied to these distances, which are
then scaled by a scalar $c_{cl_j,\ell}$, called the \emph{dual coefficient}.
The resulting values are summed over all support vectors of the class, and a
bias term $i$, called the \emph{intercept}, is added.
This leads to the following formula:

\begin{displaymath}
  i
  + \sum_{\ell \in \{1,\dots,k_1\}}\ (c_{cl_1,\ell} * f(x \otimes S_{cl_1,\ell}))
  - \sum_{\ell \in \{1,\dots,k_2\}}\ (c_{cl_2,\ell} * f(x \otimes S_{cl_2,\ell})).
\end{displaymath}
The input $x$ is classified as belonging to class $cl_1$ if the sum is positive,
and to class $cl_2$ if it is negative (the probability that it equals zero is
negligible).

For SVMs with more than two classes, we adopt the one-versus-one (\textsc{ovo})
scheme. A binary classifier is trained for each pair of classes, using dual
coefficients and intercepts specific to that pair.
(The support vectors associated with each class remain the same, although some
dual coefficients may be zero, effectively ignoring certain support vectors in a
given comparison.)
For each input, the SVM counts how many times each class is selected in these
pairwise comparisons.
For instance, with $10$ classes, the
output is a vector of $10$ integers, each between $0$ and $9$.
It should be clear that ties---where two or more classes share the highest
score---can occur, but are rare in the case of binary classification.

\begin{figure}[t!]
  \begin{center}
    \begin{tikzpicture}[ONNX/.style={fill=black!10,rounded corners},
      -|-/.style={
      to path={
          (\tikztostart) -| ($(\tikztostart)!#1!(\tikztotarget)$) |- (\tikztotarget)
          \tikztonodes
        }
      },
      -|-/.default=0.5,
      |-|/.style={
      to path={
          (\tikztostart) |- ($(\tikztostart)!#1!(\tikztotarget)$) -| (\tikztotarget)
          \tikztonodes
        }
      },
      |-|/.default=0.5,
      UNDERBRACE/.style={
          thick,
          decoration={
              brace,
              mirror,
              raise=0.5cm
            },
          decorate
        },
      ]
      \usetikzlibrary{snakes}
      \node (I) at (0, 0) {\kw{X}};
      \node (SV) at (1.5, 1) {$S$};
      \node (K) [ONNX] at (1.5, 0) {\kw{Gemm}};
      \node (Params) at (3, 1) {\begin{tabular}{c}Kernel\\params\end{tabular}};
      \node (F) [ONNX] at (3, 0) {$f$};
      \node (Duals) at (4.5, 1) {$c$};
      \node (Sum) [ONNX] at (4.5, 0) {\kw{Gemm}};
      \node (Int) at (6, 1) {$i$};
      \node (Intercept) [ONNX] at (6, 0) {\kw{Add}};
      \node (Sign) [ONNX] at (7.5, 0) {\kw{Sign}};
      \node (Scores) [ONNX] at (9, 0) {\kw{Gemm}};
      \node (Y) at (10.5, 0) {\kw{Y}};

      \draw[->] (I) to (K);
      \draw[->] (SV) to (K);
      \draw[->] (K) to (F);
      \draw[->] (Params) to (F);
      \draw[->] (F) to (Sum);
      \draw[->] (Duals) to (Sum);
      \draw[->] (Sum) to (Intercept);
      \draw[->] (Int) to (Intercept);
      \draw[->] (Intercept) to (Sign);
      \draw[->] (Sign) to (Scores);
      \draw[->] (Scores) to (Y);

      \draw [UNDERBRACE] (K.west) -- (F.east)
      node [pos=0.5,anchor=north,yshift=-0.55cm] {kernel computation};
      \draw [UNDERBRACE] (Sum.west) -- (Sign.east)
      node [pos=0.5,anchor=north,yshift=-0.55cm] {\phantom{pk}one vs. one\phantom{pk}};
      \draw [UNDERBRACE] (Scores.west) -- (Scores.east)
      node [pos=0.5,anchor=north,yshift=-0.55cm] {\phantom{pk}adding up scores\phantom{pk}};

    \end{tikzpicture}
  \end{center}
  \caption{\nir{} representation of an SVM.}%
  \label{fig:ovo}
\end{figure}

The translation of an SVM into a \nir{} is illustrated in Fig.~\ref{fig:ovo}.
The first stage computes the \emph{kernel}, which involves taking the product of
the support vectors with the input, optionally followed by the non-linear
function $f$.
The second stage performs the one-versus-one comparisons: kernel values are
multiplied by the dual coefficients, the intercept is added, and a positivity
check determines the predicted class for each pair of classes.
Finally, the last layer aggregates these pairwise scores to produce a final
score for each class.

\caisar currently supports Linear and Polynomial kernels, as well as Radial
Basis Function (RBF) kernels.


%
\subsection{The Verification Module}
\label{sec:verif}
\caisar leverages the \yyy architecture to generate verification conditions for
its supported provers.
A high-level \yml specification first undergoes an interpretation phase, during
which constants and computable expressions are evaluated and lifted to the top
level.
This is followed by a series of formula-wide transformations, handled by \yyy’s
modular infrastructure.
Prover-specific transformations may then be applied before translating the
result into the input format of the target prover.
Finally, \yyy coordinates the execution of provers and collects their results,
presenting them back to the user.

The remainder of this section describes how this general workflow is adapted to
ML programs.

\subsubsection{Interpretation of Built-Ins}
\label{sec:interpretation}
To convert a high-level \yml specification into lower-level \vnnl conditions
accepted by existing verifiers, \caisar interprets the \emph{built-ins} defined
in~\cref{sec:speclang} as follows:

\begin{itemize}
  \item \emph{Reading models and datasets.} The built-in function
        \verb|read_model| is treated as a pure function that loads an \onnx model from a
        given file path.
        Note that we do not add I/O capabilities to \yyy as \verb|read_model|
        only performs read-only access to the file system.
        Semantically, it is modeled as a mathematical function mapping valid
        \onnx file paths to their corresponding parsed models.

        Similarly, \verb|read_dataset| reads a labeled dataset from a CSV file
        and returns a vector of label-features pairs.
        Accessing this vector at a fixed index yields the corresponding data
        point.

  \item \emph{Vector computation and access.} To manipulate vectors, \verb|mapi|
        applies a user-defined function to each element of a vector whose size
        is statically known.
        The operator \verb|@@| applies a ML model---obtained via
        \verb|read_model|---to a vector of inputs and returns an output vector.
        Standard indexing \verb|v[i]| retrieves the \texttt{i}-th element of the
        vector \verb|v|.

  \item \emph{Quantification.} Bounded quantifications over vectors are
        interpreted as quantifications over their individual elements, assuming
        the vector's size is fixed and known---either explicitly provided or
        inferred from hypotheses.
        The reduction engine in \yyy maintains symbolic representations of each
        vector element, enabling the interpretation of \verb|mapi| and similar
        higher-order constructs.

\end{itemize}

\paragraph{Limitations}

While interpretation enables the expression of rich specifications, some \yml
constructs remain outside its scope.
Specifically, universal quantification over vectors is unsupported when the
vector's shape cannot be determined statically.
Since \vnnl is a fragment of the \smtl theory of Quantifier-Free Linear Real
Arithmetic (\qflra), quantifiers must be eliminated to generate valid
verification conditions.
In particular, formulas involving alternating quantifiers are not directly
supported.
\subsubsection{Embedding Specifications within Neural Networks}
\label{sec:merging}
\paragraph{Illustrative example}

Let $nn_1$ and $nn_2$ be two NNs with one (resp. two) inputs and one output
each.
Consider the following \yml expression:
\begin{equation*}
  nn_2@@(nn_1@@(x_1),x_1+\epsilon) + nn_1@@(x_0)
\end{equation*}
In this expression, $nn_1$ is applied to both $x_0$ and $x_1$; the result of
$nn_1@@(x_1)$ is used as the first input to $nn_2$, whose second input is $x_1 +
  \epsilon$.
Let $H$ be a valid \qflra{} formula defining bounds on $x_1,x_2$ and $\epsilon$.
Then, the formula

\begin{align}
  \begin{split}
    \forall x_0,x_1,\epsilon. & \ H(x_0,x_1,\epsilon) \rightarrow                                                                                       \\
                              & \overbrace{\underbrace{nn_2@@(nn_1@@}_{\text{multiple networks}}(x_1),\underbrace{x_1+\epsilon}_{\text{operation on the
      input}})}^{\text{composition of nns}}\ +\ nn_1@@(x_0) > 0
  \end{split}
  \label{eq:no_morph}
\end{align}
illustrates several limitations of the \vnnl format discussed
in~\cref{subsec:intro_vnnlib}:
\begin{enumerate}
  \item It includes a nontrivial computation on the input, $x_1+\epsilon$;
  \item It composes multiple networks, $nn_1$ and $nn_2$;
  \item It feeds the output of one network ($nn_1$) into another ($nn_2$).
\end{enumerate}

These constructs are not directly supported by \vnnl, preventing the
specification and verification of such formulas.

However, it is possible to lift these limitations by \emph{embedding} part of
the specification within a new neural network, say $nn_3$. The resulting formula
becomes:
\begin{equation}
  \forall x_0,x_1,\epsilon.\ H(x_0,x_1,\epsilon) \rightarrow nn_3@@(x_0,x_1,\epsilon) > 0
  \label{eq:morph}
\end{equation}
Here, $nn_3$ encodes the entire computation: the arithmetic on inputs, the
nested evaluation of networks, and the final comparison.
A representation of the \nir corresponding to $nn_3$ is shown in
\figurename\ref{fig:nn_combined}.
Embedding multiple networks within a single \nir allows specifications that
involve their composition, while incorporating fragments of \yml expressions
into the \nir enables more expressive specifications that capture computations
over both inputs and outputs.

The key insight is that \emph{certain \yml expressions can be encoded as \onnx
  operators}, effectively relocating parts of the specification into the neural
network itself.
This approach not only enables the expression of richer specifications that are
otherwise inexpressible in \vnnl, but also leverages the fact that \onnx
operators are more widely supported by state-of-the-art verifiers.
Combined with the interpretation mechanism introduced
in~\cref{sec:interpretation}, this strategy significantly broadens the class of
properties that can be formally specified and verified for machine learning
programs.

\DeclareRobustCommand\myNNexampleONE{\raisebox{-3pt}{\tikz \node (N11) [fill=black!10,draw,rounded corners] at (4, 1) {$nn_1$};}}
\DeclareRobustCommand\myNNexampleTWO{\raisebox{-3pt}{\tikz \node (N11) [fill=black!10,draw,rounded corners] at (4, 1) {$nn_2$};}}

\begin{figure}[h!]
  \begin{tikzpicture}
    \node (I) [fill=black!10,rounded corners] at (0,0) {\kw{Input}};
    \node (G0) [fill=black!10,rounded corners] at (2, 1.) {\kw{Gather}(0)};
    \node (G1) [fill=black!10,rounded corners] at (2, 0) {\kw{Gather}(1)};
    \node (G2) [fill=black!10,rounded corners] at (2, -1.) {\kw{Gather}(2)};
    \node (N11) [fill=black!10,draw,rounded corners] at (4, 1) {$nn_1$};
    \node (N12) [fill=black!10,draw,rounded corners] at (4, 0) {$nn_1$};
    \node (Add1) [rectangle 1x1x3x1, fill=black!10,rounded corners] at (4, -1) {\kw{Add}};
    \node (Concat) [rectangle 1x1x3x1,fill=black!10,rounded corners] at (6, -0.5) {\kw{Concat}};
    \node (N2) [fill=black!10,draw,rounded corners] at (8, -0.5) {$nn_2$};
    \node (Add2) [rectangle 1x1x3x1,fill=black!10,rounded corners] at (10, 0) {\kw{Add}};
    \draw[->] (I.east) -- ($(I.east)!0.5!(G1.west)$) coordinate (M1) -- (G1.west);
    \draw[->] (M1) -- (M1 |- G0.west) -- (G0.west);
    \draw[->] (M1) -- (M1 |- G2.west) -- (G2.west);
    \draw[->] (G0) -- (N11);
    \draw[->] (G1) -- (N12);
    \draw[->] (G1.east) -- ($(G1.east)!0.5!(N12.west)$) coordinate (M2) -- (N12.west);
    \draw[->] (M2) -- (M2 |- Add1.l2) -- (Add1.l2);
    \draw[->] (G2.east |- Add1.l1) -- (Add1.l1);
    \draw[->] (N12.east) -- ($(N12.east)!0.5!(N12.east -| Concat.l2)$) coordinate (M3) -- (M3 |- Concat.l2) -- (Concat.l2);
    \draw[->] (Add1.east) -- (Add1.east -| M3) -- (M3 |- Concat.l1) -- (Concat.l1);
    \draw[->] (Concat) -- (N2);
    \draw[->] (N2.east) -- ($(N2.east)!0.5!(N2.east -| Add2.l2)$) coordinate (M3) -- (M3 |- Add2.l1) -- (Add2.l1);
    \draw[->] (N11.east) -- (N11.east -| M3) -- (M3 |- Add2.l2) -- (Add2.l2);
    \draw[->] (Add2.east) -- ++(0.25,0);
  \end{tikzpicture}
  \caption{\kw{Gather}(0) (resp. \kw{Gather}(1)) extracts $x_0$ (resp. $x_1$)
    and \kw{Gather}(2) extracts $\epsilon$ from the \kw{Input} node.
    The first \kw{Add} node computes $x_1+\epsilon$.
    Nodes \myNNexampleONE~\myNNexampleTWO~represents
    inlined $nn_1$ and $nn_2$ control flows.
    Finally, \kw{Concat} prepares the inputs of $nn_2$.
  }
  \label{fig:nn_combined}
\end{figure}

\subsubsection{Integrating \yml Terms into \nir}

Given a \yml formula $\phi(S,\mathcal{N},P,Q)$, with $\mathcal{N}$ a
non-empty set of NNs, $S$ a set of universally quantified variables that
represents both inputs and outputs of NNs, $P$ a precondition and $Q$ a
postcondition, the goal is to write a new formula
$\phi{'}(S{'},\mathcal{N}{'},P{'},Q{'})$, equivalent to $\phi$, with
$\mathcal{N}{'}$ consisting of a single NN.

To do so, we perform a structural analysis of $\phi$, heavily relying on
\tool{OCaml} pattern-matching capabilities. First, \yml terms $t$ describing
operations not supported by \vnnl are matched. Then, subterms $t_i$s are
selected top-down. If $t_i$ can be translated into \nir nodes, then \caisar
generates a new \nir with the properly inserted nodes, removes the computation
from the \yyy formula and replaces the proper symbols.
A table representing the correspondance between \yml terms and \nir operators is
shown in \figurename\ref{fig:ConvWhy3ONNX}.

\begin{figure}
  \begin{align*}
    \mapyyytonir{e_1 + e_2}_M           & = \kw{Add}(\mapyyytonir{e_1}_M,\mapyyytonir{e_2}_M)                                            \\
    \mapyyytonir{e_1 - e_2}_M           & = \kw{Sub}(\mapyyytonir{e_1}_M,\mapyyytonir{e_2}_M)                                            \\
    \mapyyytonir{e_1 * e_2}_M           & = \kw{Mul}(\mapyyytonir{e_1}_M,\mapyyytonir{e_2}_M)                                            \\
    \mapyyytonir{e_1 / e_2}_M           & = \kw{Div}(\mapyyytonir{e_1}_M,\mapyyytonir{e_2}_M)                                            \\
    \mapyyytonir{- e_1}_M               & = \kw{Mul}(\mapyyytonir{-1.}_M,\mapyyytonir{e_2}_M)                                            \\
    \mapyyytonir{\text{\nonterm{id}}}_M & = \kw{Gather}(\kw{Input},M(\text{\nonterm{id}}))                                               \\
    \mapyyytonir{nn@@(e_1,\dots,e_n)[c]}_M
                                        & = \kw{Gather}(\texttt{Apply}(nn,\kw{Concat}(\mapyyytonir{e_1}_M,\dots,\mapyyytonir{e_n}_M)),c) \\
    \texttt{Apply}(nn,g)                & = Parser(nn)[\kw{Input} \leftarrow g]
  \end{align*}
  \caption{Mapping between \yml expressions and \nir nodes.
  All nodes are one-dimensional tensors.
  A transformation $\mapyyytonir{\cdot}_M$ is parametrized by a
  mapping of input symbols \nonterm{id} $\in S$ towards \nir input indices.
  For NNs $nn_1$ and $nn_2$, $nn_1[\kw{Input} \leftarrow nn_2]$
  builds a copy of $nn_1$ where the input $g$
  is replaced by $nn_2$.}
  \label{fig:ConvWhy3ONNX}
\end{figure}

\paragraph{Limitations}

In principle, our approach applies to any \yml expression that can be encoded as
\nir nodes.
For instance, \onnx includes an \kw{If} operator for representing conditionals.
However, such constructs are rarely supported by existing verifiers, often
requiring formula splitting and thereby increasing the number of generated
verification conditions.
Consequently, we restrict ourselves to simple arithmetic operations expressible
within the \qflra{} fragment.

Moreover, the translation of \yml formulas relies on a flattened representation
of the inputs, requiring the insertion of multiple one-dimensional \nir nodes.
Since NN verifiers are typically optimized for high-dimensional tensor
computations, this flattening introduces overhead, as empirically evaluated
in~\cref{sec:acas}.

More broadly, the question of how much of the specification should be embedded
within the neural network itself remains open.
There is an inherent trade-off between, on the one hand, the expressiveness and
conciseness afforded to the user, and on the other hand, the tractability of the
resulting verification conditions.
\subsubsection{Provers}
\label{sec:provers}
\caisar supports the following ML-specific verification tools:
\abcrown~\cite{wang2021beta},
\pyrat~\cite{lemesle2024pyrat},
\marabou~\cite{katz2019marabou,wu2024marabou},
\nnenum~\cite{bak2021nnenum},
\aimos~\cite{lemesle2023aimos},
\saver~\cite{saver}.
Thanks to the \yyy platform and the \nir, it supports translating specifications
into the \smtl format, compliant with \tool{CVC5}~\cite{barbosa2022cvc5}, \tool{Z3}~\cite{demoura2008z3} and
\tool{Alt-Ergo}~\cite{conchon2018alt} provers. It also supports gradient-boosted
trees generated with the \tool{XGBoost} library~\cite{chen2016xgboost}.
Each prover is automatically detected by \caisar if installed. An
experimental distribution of \caisar using the \tool{Nix} declarative package manager
bundles \caisar, \marabou and Python-based provers.
By default, \caisar brings an appropriate configuration for each prover.
It is possible for the user to define alternative prover configurations in the
\texttt{caisar-detection-data.conf} file. The user only needs to specify the
desired command-line options in this file, a name for the custom configuration,
and run \caisar by specifying said name.

\section{Use Cases and Evaluation}
\label{sec:eval}
\caisar provides a publicly available
manual\footnote{\url{https://caisar-platform.com/documentation}},
including classical examples from the \acas benchmark and local robustness on
\textsc{MNIST}.
It was also used in industrial contexts~\cite{cea2024caisar},
for instance in the DeepGreen\footnote{\url{https://deepgreen.ai}}
project and
Confiance.ai\footnote{\url{https://confiance.ai}} program.
In the following, we also illustrate \caisar
capabilities to specify properties beyond local robustness properties.

\subsection{Formal Verification of \acas with Unnormalized Inputs}
\label{sec:acas}
We consider the classical \acas benchmarks, originally presented
in~\cite{katz2017reluplex} with a subtle, but important difference. The original
paper presents the benchmark with unnormalized values. For instance, property
$\phi_1$ is written as follows:
\begin{equation*}
  \begin{split}
    \mathcal{P}(x) =
    x[\rho] & \geq 55947.691\  \land\  x[vown] \geq 1145\ \land\ x[vint] \leq 60 \\
            & \implies  \mathcal{Q}(y) = y[coc] < 1500
  \end{split}
\end{equation*}
However, the actual property specifications encoded in \marabou exhibit
normalized values\footnote{Values available on the
  \href{https://github.com/NeuralNetworkVerification/Marabou/blob/215828c64e624be7917e69e4e873c746d8df25a2/resources/properties/acas_property_1.txt}{\marabou
    GitHub repository.}}.
As neural networks require normalized inputs, this is expected;
however it creates a gap between the specification expressed
in terms of expert-domain knowledge
and the actual values checked during verification. This \emph{embedding gap}
was identified as a major obstacle~\cite{cordeiro2025neural} to actionable
formal specification.

Fortunately, specifying the normalization process inside the \yml
specification will produce an \onnx file \emph{that normalizes its input
  according to the specification}.
The network will then compute normalized results,
which can then be returned to the original output space. As
long as the normalization and denormalization processes are well-specified, the
specification can be expressed on the original input and output.
\figurename\ref{fig:acas_unnormalized} provides a \yml excerpt of such 
specification.

\begin{figure}[htp]
  \begin{lstlisting}
  let function normalize_t i mean range = (i .- mean) ./ range
  let function denormalize_t i mean range = (i .* range) .+ mean
  let function normalize_input i = Vector.mapi i normalize_by_index
  
  let function denormalize_output_t o =
    denormalize_t o
      (7.51888402010059753166615337249822914600372314453125:t)
        (373.9499200000000200816430151462554931640625:t)
  
  let runP1 (i: input) : t
    requires { has_length i 5 /\ valid_input i }
    requires { intruder_distant_and_slow i }
    ensures { result .<= (1500.0:t) }  =
      let j = normalize_input i in
      let o = (nn @@ j)[clear_of_conflict] in
      (denormalize_output_t o)\end{lstlisting}
  \caption{A \yml specification for the unnormalized \acas $\phi_1$ property.}
  \label{fig:acas_unnormalized}
\end{figure}

In \tablename~\ref{tab:provers_acas_norm}, we report runtimes on the original
(normalized) \acas properties as well as their unnormalized counterparts.
For \pyrat, our graph-editing technique preserves the verification outcome for
all properties. However, the additional overhead is significant enough to cause
many timeouts within the allocated runtime budget.
For \nnenum, results remain consistent except for occasional crashes, which stem
from a known bug in its \onnx parsing module.
By contrast, \maraboupy, the Python wrapper of \marabou, exhibits incorrect
behavior: it reports counterexamples for properties $\phi_4$, $\phi_5$, and
$\phi_9$, where none should exist according to the original specification. Even
more concerning, \maraboupy finds a counterexample for the normalized version of
$\phi_{10}$, in direct contradiction with \marabou itself, which correctly
verifies the property.
During our experiments, \maraboupy raised degradation warnings about
floating-point arithmetic on $\phi_{10}$, which may explain this discrepancy. In
particular, \caisar outputs specifications using \texttt{IEEE 754}
double-precision floating-point arithmetic, whereas the \acas networks---and the
instances generated from them---operate exclusively in single precision.
Further investigation revealed an inconsistency between the plain \marabou
solver, which supports only a limited set of \onnx operators, and its
\maraboupy, which supports a broader set. This issue was reported to the
\marabou
developers\footnote{\url{https://github.com/NeuralNetworkVerification/Marabou/issues/882}}.

Nevertheless, this example displays how \caisar allows users to compare multiple
provers on the same specification; further helping to build a case on the
correctness of a program.

\begin{table}[t!]
  \caption{\acas benchmarks on selected verifiers,
    averaged over three runs. The
    string next to the property indicates the
    \tool{Reluplex}~\cite{katz2017reluplex} neural network it was verified
    against. Columns $T_n$ (resp. $T_u$) report the time (in seconds) required
    by verifiers to answer on the original (resp. the unnormalized) queries,
    while columns $A_n$ (resp. $A_u$) report the actual answers on the original
    (resp. the unnormalized) queries. A \cmark{} indicates that a property is
    satisfied, a \xmark{} that a counterexample has been found, a
    \VarClock\xspace that a timeout is reached, and a $?$ that a runtime crash
    occurred. Failures of \marabou are expected on unnormalized properties, as
    the \kw{Concat} operator, mandatory for our approach, is not supported.}
  \label{tab:provers_acas_norm}
  \begin{center}
    \resizebox{\textwidth}{!}{
      \begin{tabular}{c||c|c|c|c||c|c|c|c||c|c|c|c||c|c|c|c}
  \textbf{Property} & \multicolumn{4}{|c||}{\textbf{\marabou}} & \multicolumn{4}{|c||}{\textbf{\maraboupy}} & \multicolumn{4}{|c||}{\textbf{\pyrat}} & \multicolumn{4}{|c}{\textbf{\nnenum}}                                                                                                                                     \\ \hline
                    & $T_n$                                    & $A_n$                                      & $T_u$                                  & $A_u$                                 & $T_n$  & $A_n$       & $T_u$   & $A_u$       & $T_n$  & $A_n$       & $T_u$  & $A_u$       & $T_n$  & $A_n$    & $T_u$ & $A_u$    \\ \hline
  $\phi_{1}$        & 3.00                                     & (?)                                        & 3.00                                   & (?)                                   & 5.00   & (\cmark)    & 243.00  & (\VarClock) & 8.00   & (\cmark)    & 11.00  & (\cmark)    & 4.00   & (\cmark) & 4.00  & (\cmark) \\
  $\phi_{2}$        & 37.00                                    & (\cmark)                                   & 3.00                                   & (?)                                   & 26.00  & (\cmark)    & 243.00  & (\VarClock) & 19.00  & (\cmark)    & 38.00  & (\cmark)    & 4.00   & (\cmark) & 6.00  & (?)      \\
  $\phi_{3}$        & 243.00                                   & (\VarClock)                                & 5.00                                   & (?)                                   & 243.00 & (\VarClock) & 243.00  & (\VarClock) & 246.00 & (\VarClock) & 246.00 & (\VarClock) & 4.00   & (\cmark) & 4.00  & (\cmark) \\
  $\phi_{4}$        & 44.00                                    & (\cmark)                                   & 5.00                                   & (?)                                   & 36.00  & (\cmark)    & 4.00    & (\xmark)    & 25.00  & (\cmark)    & 246.00 & (\VarClock) & 4.00   & (\cmark) & 4.00  & (\cmark) \\
  $\phi_{5}$        & 102.00                                   & (\cmark)                                   & 5.00                                   & (?)                                   & 93.00  & (\cmark)    & 5.00    & (\xmark)    & 246.00 & (\VarClock) & 246.00 & (\VarClock) & 4.00   & (\cmark) & 5.00  & (\cmark) \\
  $\phi_{6}$        & 558.00                                   & (\cmark)                                   & 5.00                                   & (?)                                   & 566.00 & (\cmark)    & 1925.00 & (\VarClock) & 156.00 & (\cmark)    & 426.00 & (\VarClock) & 7.00   & (\cmark) & 13.00 & (?)      \\
  $\phi_{7}$        & 485.00                                   & (\VarClock)                                & 5.00                                   & (?)                                   & 484.00 & (\VarClock) & 484.00  & (\VarClock) & 246.00 & (\VarClock) & 246.00 & (\VarClock) & 119.00 & (\xmark) & 4.00  & (?)      \\
  $\phi_{8}$        & 485.00                                   & (\xmark)                                   & 5.00                                   & (?)                                   & 8.00   & (\xmark)    & 248.00  & (\VarClock) & 246.00 & (\VarClock) & 246.00 & (\VarClock) & 4.00   & (\xmark) & 4.00  & (\xmark) \\
  $\phi_{9}$        & 182.00                                   & (\cmark)                                   & 5.00                                   & (?)                                   & 222.00 & (\cmark)    & 5.00    & (\xmark)    & 61.00  & (\cmark)    & 246.00 & (\VarClock) & 6.00   & (\cmark) & 9.00  & (\cmark) \\
  $\phi_{10}$       & 83.00                                    & (\cmark)                                   & 3.00                                   & (?)                                   & 151.00 & (\xmark)    & 245.00  & (\xmark)    & 13.00  & (\cmark)    & 246.00 & (\VarClock) & 4.00   & (\cmark) & 5.00  & (\cmark) \\
\end{tabular}
    }
  \end{center}
\end{table}

\subsection{On the Correctness of SVM Parsing}
\label{sec:nnsvm}
In the context of one-versus-one (\textsc{ovo}) classification, consider an SVM
$f$, a point $x \in \mathbb{R}^d$, a class $cl \in \mathbb{N}$, and a
perturbation bound $\epsilon \in \mathbb{R}$. The \emph{local robustness}
property requires that for all $x'\in \mathbb{R}^d$ such that $\|x - x'\| \leq
  \epsilon$, the class maximizing the output of $f$ remains $cl$.

For evaluation, we considered the model \texttt{linear-1k} from the
\textsc{MNIST} domain provided by the developers of \saver{}\footnote{The model
  is available at
  \url{https://github.com/abstract-machine-learning/data-collection/tree/master/domains/mnist/models/svm/linear-1k.dat.zip}}
with the associated dataset\footnote{The dataset is available at
  \url{https://github.com/abstract-machine-learning/data-collection/blob/master/domains/mnist/datasets/test-set.csv.zip}}.
We provided the robustness property to \caisar{} using the translation from SVM
to \nir{} described in~\cref{sec:nir}, and used \pyrat{} as verifier.
We then compared the results with those produced by \saver.

\begin{table}
  \centering
  \caption{Runtime comparison between \saver{} and \caisar{}.
    \emph{Valid} indicates that robustness holds for the instance.
    The two solvers agree on the result.}\label{tab::svm}
  \begin{tabular}{
    >{\centering\hspace{0pt}}p{.2\linewidth} |
    >{\centering\hspace{0pt}}p{.2\linewidth} |
    >{\centering\hspace{0pt}}p{.2\linewidth} |
    >{\centering\hspace{0pt}}p{.3\linewidth}}
    \textbf{Instance} & \textbf{Result} & \saver{} (ms) & \pyrat{} via \caisar{} (ms) \tabularnewline\hline
    \ 0               & Valid           & 35            & 19,101\tabularnewline{}
    1                 & Invalid         & 49            & 20,740\tabularnewline{}
    2                 & Valid           & 53            & 18,552\tabularnewline{}
    3                 & Valid           & 41            & 22,001\tabularnewline{}
    4                 & Valid           & 45            & 20,464\tabularnewline{}
    5                 & Valid           & 47            & 19,300\tabularnewline{}
    6                 & Valid           & 43            & 20,546\tabularnewline{}
    7                 & Valid           & 48            & 19,802\tabularnewline{}
    8                 & Invalid         & 43            & 20,111\tabularnewline{}
    9                 & Valid           & 42            & 20,957
  \end{tabular}
\end{table}

The results are presented in Table~\ref{tab::svm}.
The answers produced by \caisar{} are consistent with those of \saver{}, but
obtained at a runtime approximately two orders of magnitude higher,
independently of whether the robustness property holds.
It should be noted, however, that \saver{} was designed specifically for the
verification of robustness properties and cannot be applied to other classes of
properties.
In addition, \saver{} no longer appears to be under active development, casting
doubts on its long-term usability.
By contrast, \caisar{} is designed as a general specification and verification
framework, and ongoing work aims to improve its performance on this class of
problems.

\subsection{Composition of Neural Networks}
\label{sec:composition}
We evaluated \caisar{} on two illustrative specifications concerning multiple
NNs: a property about the output of the sequential composition of two NNs
(\cref{fig:specexamplesequencing}) and the comparison of outputs between two NNs
(\cref{fig:specexample}).
Both serve as representative examples of practical verification tasks, such as
quantifying the effect of quantization or ensuring correctness of composed
models.

For the first specification (see \cref{fig:specexamplesequencing}), which
involves sequential composition of NNs, we evaluated \caisar with \maraboupy,
\nnenum and \pyrat as verifiers.
Within a runtime budget of one hour, \maraboupy could not prove the property
(most likely by finding a counterexample), while both \nnenum and \pyrat timed
out (see \cref{tab:seqres}).
It is worth noting that \marabou{} cannot be used as a backend prover for this
property, since it does not support all \onnx operators required to encode the
sequential combination of the two networks.

\begin{table}[h]
  \centering
  \caption{Verification results of the sequential composition property of
    \cref{fig:specexamplesequencing}.}
  \label{tab:seqres}
  \begin{tabular}{
    >{\centering\hspace{0pt}}p{.2\linewidth} |
    >{\centering\hspace{0pt}}p{.2\linewidth}}
    \textbf{Prover} & \textbf{Result} \tabularnewline\hline
    \maraboupy      & Invalid           \tabularnewline{}
    \nnenum         & Timeout         \tabularnewline{}
    \pyrat          & Timeout
  \end{tabular}
\end{table}

For the second specification (see \cref{fig:specexample}), the results are not
revealing, as all provers ultimately reach timeouts without producing a
conclusive answer.

Overall, these experiments demonstrate that \caisar{} can express non-trivial
properties beyond local robustness, dispatching their verification to multiple
backend tools, though performance and conclusiveness remain highly dependent on
the underlying prover.

\section{Conclusion}
\label{sec:futurework}
Historically, the formal verification of human-written software evolved from tools able to verify simple properties to large-scale platforms (\eg \yyy) capable of handling higher-level properties, decomposing them into smaller properties, and dispatching them to back-end solvers and analyzers.
The AI field, and in particular ML, already advanced to the first stage of such evolution.
We present the extensible, open-source \caisar platform, that aims to embody the next stage of that similar evolution, with emphasis on an expressive language where higher-level properties can be described, as well as the compilation of these properties into inputs that existing solvers can handle.
We also expanded the range of what is possible to specify with the clear goal of encouraging the community to target such specifications in the next versions of their solvers.
The authors are keen for any request for collaboration or special needs, such as handling particular model architectures or properties.

Of particular interest is the integration of confidence-based properties, as well as expanding the set of supported tools. We are also actively developing heuristics to inform the validation process, particularly through past accumulated knowledge of successful runs (\eg on use-cases involving X type of properties on Y type of model with Z activation functions, provers A gave the best results with configuration C, therefore a suggested strategy is to start with A on use-cases with similar characteristics to X, Y and Z). An ongoing collaboration is focused on bridging the validation process with requirement-based engineering.

\caisar is made available online, with clear documentation as well as tutorials that can be used in teaching classes to infuse the paramount importance of safety in the minds of future generations of ML developers.
\caisar's existence goes beyond this sole publication: it has the durable support of CEA-List and a dedicated developing team, giving the platform the stable future necessary for long-term support.


\subsubsection*{Acknowledgements}
This work was partially supported by the SAIF project (ANR-23-PEIA-0006) and the
DeepGreen project (ANR-23-DEGR-0001) funded by the ``France 2030'' government
investment plan managed by the French National Research Agency (ANR).

The authors thank Serge Durand for his help with \marabou and \pyrat, and
Augustin Lemesle and Julien Lehmann for fruitful discussions.


%
%
\bibliographystyle{splncs04}
\bibliography{laiser.bib}

\begin{thebibliography}{10}
\providecommand{\url}[1]{\texttt{#1}}
\providecommand{\urlprefix}{URL }
\providecommand{\doi}[1]{https://doi.org/#1}

\bibitem{alberti2022caisar}
Alberti, M., Bobot, F., Chihani, Z., Girard-Satabin, J., Lemesle, A.: {CAISAR:
  A platform for Characterizing Artificial Intelligence Safety and Robustness}.
  In: {AISafety}. CEUR-Workshop Proceedings, Vienne, Austria (Jul 2022),
  \url{https://hal.archives-ouvertes.fr/hal-03687211}

\bibitem{athavale2024verifying}
Athavale, A., Bartocci, E., Christakis, M., Maffei, M., Nickovic, D.,
  Weissenbacher, G.: Verifying global two-safety properties in neural networks
  with confidence. In: Gurfinkel, A., Ganesh, V. (eds.) Computer Aided
  Verification. pp. 329--351. Springer Nature Switzerland, Cham (Jun 2024).
  \doi{10.48550/arXiv.2405.14400}, \url{http://arxiv.org/abs/2405.14400}, zSCC:
  0000000 arXiv:2405.14400 [cs] type: article

\bibitem{audemard2024computation}
Audemard, G., Lagniez, J.M., Marquis, P.: On the Computation of Contrastive
  Explanations for Boosted Regression Trees. IOS Press (Oct 2024).
  \doi{10.3233/faia240600}

\bibitem{bak2021nnenum}
Bak, S.: Nnenum: {{Verification}} of {{ReLU Neural Networks}} with {{Optimized
  Abstraction Refinement}}. In: Dutle, A., Moscato, M.M., Titolo, L.,
  Mu{\~n}oz, C.A., Perez, I. (eds.) {{NASA Formal Methods}}. pp. 19--36.
  Lecture {{Notes}} in {{Computer Science}}, {Springer International
  Publishing}, {Cham} (2021). \doi{10.1007/978-3-030-76384-8_2}

\bibitem{balan2017lipschitz}
Balan, R., Singh, M., Zou, D.: Lipschitz properties for deep convolutional
  networks (2017). \doi{10.48550/ARXIV.1701.05217}

\bibitem{balunovic2019certifying}
Balunovic, M., Baader, M., Singh, G., Gehr, T., Vechev, M.: Certifying
  {Geometric} {Robustness} of {Neural} {Networks}. In: Advances in {Neural}
  {Information} {Processing} {Systems}. vol.~32. Curran Associates, Inc.
  (2019),
  \url{https://papers.nips.cc/paper/2019/hash/f7fa6aca028e7ff4ef62d75ed025fe76-Abstract.html}

\bibitem{barbosa2022cvc5}
Barbosa, H., Barrett, C.W., Brain, M., Kremer, G., Lachnitt, H., Mann, M.,
  Mohamed, A., Mohamed, M., Niemetz, A., N{\"{o}}tzli, A., Ozdemir, A.,
  Preiner, M., Reynolds, A., Sheng, Y., Tinelli, C., Zohar, Y.: cvc5: {A}
  versatile and industrial-strength {SMT} solver. In: Fisman, D., Rosu, G.
  (eds.) Tools and Algorithms for the Construction and Analysis of Systems -
  28th International Conference, {TACAS} 2022, Held as Part of the European
  Joint Conferences on Theory and Practice of Software, {ETAPS} 2022, Munich,
  Germany, April 2-7, 2022, Proceedings, Part {I}. Lecture Notes in Computer
  Science, vol. 13243, pp. 415--442. Springer (2022).
  \doi{10.1007/978-3-030-99524-9\_24},
  \url{https://doi.org/10.1007/978-3-030-99524-9\_24}

\bibitem{bassan2023towards}
Bassan, S., Katz, G.: Towards Formal XAI: Formally Approximate Minimal
  Explanations of Neural Networks, pp. 187--207. Springer Nature Switzerland
  (2023). \doi{10.1007/978-3-031-30823-9_10}

\bibitem{boetius2023verifying}
Boetius, D., Leue, S.: Verifying global neural network specifications using
  hyperproperties (2023)

\bibitem{brix2024fifth}
Brix, C., Bak, S., Johnson, T.T., Wu, H.: The fifth international verification
  of neural networks competition (vnn-comp 2024): Summary and results (2024).
  \doi{10.48550/ARXIV.2412.19985}, \url{https://arxiv.org/abs/2412.19985}

\bibitem{brix2023fourth}
Brix, C., Bak, S., Liu, C., Johnson, T.T.: The fourth international
  verification of neural networks competition (vnn-comp 2023): Summary and
  results (2023)

\bibitem{brix2023first}
Brix, C., Müller, M.N., Bak, S., Johnson, T.T., Liu, C.: First three years of
  the international verification of neural networks competition (vnn-comp)
  (2023)

\bibitem{casadio2022neural}
Casadio, M., Komendantskaya, E., Daggitt, M.L., Kokke, W., Katz, G., Amir, G.,
  Refaeli, I.: Neural network robustness as a verification property: a
  principled case study. In: International Conference on Computer Aided
  Verification. pp. 219--231. Springer (2022)

\bibitem{cea2024caisar}
CEA: Caisar (May 2024), \url{https://catalog.confiance.ai/records/px81f-r1p78}

\bibitem{chareton2021automated}
Chareton, C., Bardin, S., Bobot, F., Perrelle, V., Valiron, B.: An Automated
  Deductive Verification Framework for Circuit-building Quantum Programs, pp.
  148--177. Springer International Publishing (2021).
  \doi{10.1007/978-3-030-72019-3_6}

\bibitem{chen2016xgboost}
Chen, T., Guestrin, C.: {XGBoost}: A scalable tree boosting system. In:
  Proceedings of the 22nd ACM SIGKDD International Conference on Knowledge
  Discovery and Data Mining. pp. 785--794. KDD '16, ACM, New York, NY, USA
  (2016). \doi{10.1145/2939672.2939785},
  \url{http://doi.acm.org/10.1145/2939672.2939785}

\bibitem{conchon2018alt}
Conchon, S., Coquereau, A., Iguernlala, M., Mebsout, A.: {Alt-Ergo 2.2}. In:
  {SMT Workshop: International Workshop on Satisfiability Modulo Theories}.
  Oxford, United Kingdom (Jul 2018), \url{https://hal.inria.fr/hal-01960203}

\bibitem{conchon2007designing}
Conchon, S., Filli\^atre, J.C., Signoles, J.: {Designing a Generic Graph
  Library using {ML} Functors}. In: Moraz\'an, M.T., Nilsson, H. (eds.) The
  Eighth Symposium on Trends in Functional Programming. vol.
  TR-SHU-CS-2007-04-1, pp. XII/1--13. Seton Hall University, New York, USA (Apr
  2007), \url{https://usr.lmf.cnrs.fr/~jcf/publis/ocamlgraph-tfp07.ps}

\bibitem{cordeiro2025neural}
Cordeiro, L.C., Daggitt, M.L., Girard-Satabin, J., Isac, O., Johnson, T.T.,
  Katz, G., Komendantskaya, E., Lemesle, A., Manino, E., Šinkarovs, A., Wu,
  H.: Neural network verification is a programming language challenge. In:
  Vafeiadis, V. (ed.) Programming Languages and Systems. pp. 206--235. Springer
  Nature Switzerland (2025). \doi{10.1007/978-3-031-91118-7_9}

\bibitem{cristianini2008support}
Cristianini, N., Ricci, E.: Support Vector Machines, pp. 928--932. Springer US,
  Boston, MA (2008). \doi{10.1007/978-0-387-30162-4_415},
  \url{https://doi.org/10.1007/978-0-387-30162-4_415}

\bibitem{daggitt2022vehicle}
Daggitt, M.L., Kokke, W., Atkey, R., Arnaboldi, L., Komendantskya, E.: Vehicle:
  Interfacing neural network verifiers with interactive theorem provers (2022).
  \doi{10.48550/ARXIV.2202.05207}, \url{https://arxiv.org/abs/2202.05207}

\bibitem{daggitt2024vehicle}
Daggitt, M.L., Kokke, W., Atkey, R., Slusarz, N., Arnaboldi, L.,
  Komendantskaya, E.: Vehicle: Bridging the embedding gap in the verification
  of neuro-symbolic programs (2024). \doi{10.48550/ARXIV.2401.06379}

\bibitem{demoura2008z3}
{de Moura}, L., Bj{\o}rner, N.: Z3: {{An Efficient SMT Solver}}. In:
  Ramakrishnan, C.R., Rehof, J. (eds.) Tools and {{Algorithms}} for the
  {{Construction}} and {{Analysis}} of {{Systems}}. pp. 337--340. Lecture
  {{Notes}} in {{Computer Science}}, {Springer}, {Berlin, Heidelberg} (2008).
  \doi{10.1007/978-3-540-78800-3_24}

\bibitem{durand2022reciph}
Durand, S., Lemesle, A., Chihani, Z., Urban, C., Terrier, F.: {ReCIPH:
  Relational Coefficients for Input Partitioning Heuristic}. WFVML 2022  (Jul
  2022), \url{https://inria.hal.science/hal-03926281}, poster

\bibitem{filliatre2013why3}
Filli{\^a}tre, J.C., Paskevich, A.: Why3 - {{Where Programs Meet Provers}}. In:
  Felleisen, M., Gardner, P. (eds.) Programming {{Languages}} and {{Systems}}.
  pp. 125--128. Lecture {{Notes}} in {{Computer Science}}, {Springer}, {Berlin,
  Heidelberg} (2013). \doi{10.1007/978-3-642-37036-6_8}

\bibitem{fischer2019dl2}
Fischer, M., Balunovic, M., Drachsler-Cohen, D., Gehr, T., Zhang, C., Vechev,
  M.: {DL}2: Training and querying neural networks with logic. In: Chaudhuri,
  K., Salakhutdinov, R. (eds.) Proceedings of the 36th International Conference
  on Machine Learning. Proceedings of Machine Learning Research, vol.~97, pp.
  1931--1941. PMLR (09--15 Jun 2019),
  \url{https://proceedings.mlr.press/v97/fischer19a.html}

\bibitem{girard-satabin2020camus}
Girard-Satabin, J., Charpiat, G., Chihani, Z., Schoenauer, M.: {CAMUS: A
  Framework to Build Formal Specifications for Deep Perception Systems Using
  Simulators}. In: {ECAI 2020 - 24th European Conference on Artificial
  Intelligence}. Santiago de Compostela, Spain (Jun 2020),
  \url{https://hal.inria.fr/hal-02440520}

\bibitem{katz2017reluplex}
Katz, G., Barrett, C., Dill, D.L., Julian, K., Kochenderfer, M.J.: Reluplex: An
  Efficient SMT Solver for Verifying Deep Neural Networks, pp. 97--117.
  Springer International Publishing (2017). \doi{10.1007/978-3-319-63387-9_5}

\bibitem{katz2019marabou}
Katz, G., Huang, D.A., Ibeling, D., Julian, K., Lazarus, C., Lim, R., Shah, P.,
  Thakoor, S., Wu, H., Zeljić, A., Dill, D.L., Kochenderfer, M.J., Barrett,
  C.: The {Marabou} {Framework} for {Verification} and {Analysis} of {Deep}
  {Neural} {Networks}. In: Dillig, I., Tasiran, S. (eds.) Computer {Aided}
  {Verification}, vol. 11561, pp. 443--452. Springer International Publishing,
  Cham (2019). \doi{10.1007/978-3-030-25540-4_26}

\bibitem{koenig2024critically}
K{{\"o}}nig, M., Bosman, A.W., Hoos, H.H., van Rijn, J.N.: Critically assessing
  the state of the art in neural network verification. Journal of Machine
  Learning Research  \textbf{25}(12),  1--53 (2024),
  \url{http://jmlr.org/papers/v25/23-0119.html}

\bibitem{lemesle2024pyrat}
Lemesle, A., Lehmann, J., Gall, T.L.: Neural network verification with pyrat
  (2024). \doi{10.48550/ARXIV.2410.23903}

\bibitem{lemesle2023aimos}
Lemesle, A., Varasse, A., Chihani, Z., Tachet, D.: {AIMOS: Metamorphic Testing
  of AI - AnIndustrial Application}. WAISE 2023  (2023)

\bibitem{pulina2011never}
Pulina, L., Tacchella, A.: Never: a tool for artificial neural networks
  verification. Annals of Mathematics and Artificial Intelligence
  \textbf{62}(3–4),  403–425 (jul 2011). \doi{10.1007/s10472-011-9243-0},
  \url{https://doi.org/10.1007/s10472-011-9243-0}

\bibitem{saver}
Ranzato, F., Zanella, M.: Robustness verification of support vector machines.
  In: Chang, B.Y.E. (ed.) Static Analysis. pp. 271--295. Springer International
  Publishing, Cham (2019)

\bibitem{singh2019single}
Singh, G., Ganvir, R., Püschel, M., Vechev, M.: Beyond the {Single} {Neuron}
  {Convex} {Barrier} for {Neural} {Network} {Certification}. In: Wallach, H.,
  Larochelle, H., Beygelzimer, A., Alché-Buc, F.d., Fox, E., Garnett, R.
  (eds.) Advances in {Neural} {Information} {Processing} {Systems} 32, pp.
  15098--15109. Curran Associates, Inc. (2019),
  \url{http://papers.nips.cc/paper/9646-beyond-the-single-neuron-convex-barrier-for-neural-network-certification.pdf}

\bibitem{trojanek2014verification}
Trojanek, P., Eder, K.: Verification and testing of mobile robot navigation
  algorithms: A case study in spark. In: 2014 IEEE/RSJ International Conference
  on Intelligent Robots and Systems. pp. 1489--1494. IEEE (Sep 2014).
  \doi{10.1109/iros.2014.6942753}

\bibitem{urban2019perfectly}
Urban, C., Christakis, M., W{\"u}stholz, V., Zhang, F.: Perfectly {{Parallel
  Fairness Certification}} of {{Neural Networks}}. arXiv:1912.02499 [cs]  (Dec
  2019)

\bibitem{urban2021review}
Urban, C., Min{\'e}, A.: A {{Review}} of {{Formal Methods}} applied to
  {{Machine Learning}}. arXiv:2104.02466 [cs]  (Apr 2021)

\bibitem{walt2014scikit}
van~der Walt, S., {S}ch\"onberger, J.L., {Nunez-Iglesias}, J., {B}oulogne, F.,
  {W}arner, J.D., {Y}ager, N., {G}ouillart, E., {Y}u, T., the scikit-image
  contributors: scikit-image: image processing in {P}ython. PeerJ  \textbf{2},
  ~e453 (6 2014). \doi{10.7717/peerj.453},
  \url{https://doi.org/10.7717/peerj.453}

\bibitem{wang2021beta}
Wang, S., Zhang, H., Xu, K., Lin, X., Jana, S., Hsieh, C.J., Kolter, J.Z.:
  Beta-{{CROWN}}: {{Efficient Bound Propagation}} with {{Per-neuron Split
  Constraints}} for {{Complete}} and {{Incomplete Neural Network Robustness
  Verification}}, \url{http://arxiv.org/abs/2103.06624}

\bibitem{wu2024marabou}
Wu, H., Isac, O., Zeljić, A., Tagomori, T., Daggitt, M., Kokke, W., Refaeli,
  I., Amir, G., Julian, K., Bassan, S., Huang, P., Lahav, O., Wu, M., Zhang,
  M., Komendantskaya, E., Katz, G., Barrett, C.: Marabou 2.0: A versatile
  formal analyzer of neural networks (2024). \doi{10.48550/ARXIV.2401.14461}

\bibitem{wu2023verix}
Wu, M., Wu, H., Barrett, C.: Verix: Towards verified explainability of deep
  neural networks (2023)

\bibitem{xie2022neuro}
Xie, X., Kersting, K., Neider, D.: Neuro-symbolic verification of deep neural
  networks. In: Raedt, L.D. (ed.) Proceedings of the Thirty-First International
  Joint Conference on Artificial Intelligence, {IJCAI-22}. pp. 3622--3628.
  International Joint Conferences on Artificial Intelligence Organization (7
  2022). \doi{10.24963/ijcai.2022/503},
  \url{https://doi.org/10.24963/ijcai.2022/503}, main Track

\end{thebibliography}
%

\end{document}